\documentclass[twocolumn,showpacs,preprintnumbers,amsmath,amssymb]{revtex4}
\usepackage{graphicx}% Include figure files
\begin{document}

\preprint{}

\title{Carbon Nanotube Based Bearing for Rotational Motions}
% Force line breaks with \\

\author{B. Bourlon, D.C. Glattli}
 \altaffiliation[Also at ]{SPEC, CEA Saclay, F-91191 Gif-sur-Yvette,
France.}
\author{ A. Bachtold}%
 \email{bachtold@lpmc.ens.fr}
\affiliation{ Laboratoire de Physique de la Mati\`{e}re
Condens\'{e}e de l'Ecole Normale Sup\'{e}rieure, 24 rue Lhomond,
75231 Paris 05,
France. }%

\author{L. Forr\'{o}}
\affiliation{ EPFL, CH-1015, Lausanne, Switzerland.}%

\date{\today}% It is always \today, today,

\begin{abstract}
We report the fabrication of a nanoelectromechanical system
consisting of a plate rotating around a multiwalled nanotube
bearing. The motion is possible thanks to the low intershell
friction. Indeed, the nanotube has been engineered so that the
sliding happens between different shells. The plate rotation is
activated electrostatically with stator electrodes. The static
friction force is estimated at \mbox{$\approx2\cdot10^{-15}$
N/\AA$^2$}.

\end{abstract}

\maketitle

Microfabricated motors have been one of the most studied
microelectromechanical systems in the 80s and 90s
\cite{Trimmer,Fan,Tai,Mehregany}. Many efforts have been invested
in improving performance characteristics such as the rotational
speed. It has been shown that the characteristics are mostly
limited by the fabrication process through the low reproduction
accuracy of the design geometry \cite{Mehregany}. For example,
microfabrication cannot realize regular enough surfaces at the
contact between the rotor and the bearing. These irregularities
are important sources of energy loss and reduce considerably the
rotational speed.

Multiwalled carbon nanotubes (MWNTs) consist of several nested
cylindrical shells. Electron transmission microscopy has shown the
high degree of perfection at the atomic level of their structure
\cite{Dresselhaus}. Their structural perfection and geometry make
MWNTs attractive as elements enabling a rotational motion for
nanoelectromechanical systems (NEMS) \cite{Roukes,Craighead}. Fig.
1a shows the operating principle of such a MWNT element where one
or more inner shells can slide with respect to some outer shells.

The utilization of MWNTs as bearings for rotational motion in NEMS
has been further motivated by recent experiments that show that
the intershell friction can be low in MWNTs \cite{Cummings,Yu}. In
Ref.\cite{Cummings} the static friction force has been
experimentally estimated to be lower than \mbox{$6\cdot10^{-15}$
N/\AA$^2$}, which gives a very low friction torque due to the thin
diameters of the shells. Easy sliding motion between shells has
also been predicted theoretically \cite{Charlier,Crespi,Saito}.
Interestingly, corrugation against sliding of incommensurate and
free of disorder shells is predicted to be extremely small
\cite{Crespi}.

We report the realization of a NEMS where a metal plate rotates
around a MWNT axle. The suspended MWNT is clamped between two
anchor pads. The MWNT has been engineered with the
electrical-breakdown technique \cite{Collins1,Collins2,Bourlon},
as shown in Fig. 1(a), to allow access to the inner shells on
which is fixed the metal plate. The metal plate is shown to rotate
due to surface tension forces of a liquid drop that evaporates on
the substrate. The rotational motion can also be activated
electrostatically with two stator electrodes. In this way, the
plate does not achieve complete rotations but can be positioned in
directions lying between the two stator electrodes.

Our rotational NEMSs are fabricated through a four-step process.
The MWNTs are synthetised by arc-discharge evaporation and
carefully purified \cite{Bonard}. From a dispersion in
dichloroethane the nanotubes are dispersed onto a 1000 nm oxidized
Si wafer. The nanotubes are imaged with atomic force microscopy
(AFM). Thick MWNTs of diameter between 15 and 25 nm  are selected
because they are mechanically more robust. Their positions are
recorded with respect to metallic marks. The nanotubes are then
connected with two electrodes using electron beam lithography
(Fig. 1(b)). These electrodes will be used both as conducting
electrodes and as anchor pads that hold the nanotube bearing. The
electrodes, which are \mbox{3 $\mu$m} wide and consist of 10 nm Cr
and 60 nm Au, have to be wide and thick enough such that they stay
horizontal and suspended when the SiO2 is etched away at the end
of the process.

Fig. 1(c) shows that several outer shells are removed in between
the electrodes in order to have access to the inner ones. These
outer shells are removed with the electrical-breakdown technique
\cite{Collins1,Collins2,Bourlon} that we review briefly in this
paragraph. The bias voltage applied on the tube is continuously
increased while the current is recorded. When a sharp step in the
current is detected the bias is quickly reduced to 0. This sharp
step is usually a current reduction of \mbox{$\sim$20 $\mu$A}
which corresponds to the oxidation of the outermost shell.
Importantly, the shell is then removed along with most of the tube
portion that is not covered by the electrodes. This leaves plenty
of room for access to the inner shells.

The electrical-breakdown technique is repeated until several
shells are removed to obtain a sliding motion with a lower
friction. The sliding will occur through a self-selection process
between the shells with the most perfect surfaces that offer the
least resistance to motion. However, we cannot repeat this process
until we are left with one or a very few shells as the structure
is not robust enough to sustain the drying step at the end of the
process. We have found that the samples with inner rotating shells
of diameter around 15 nm work well.

\begin{figure}
\includegraphics{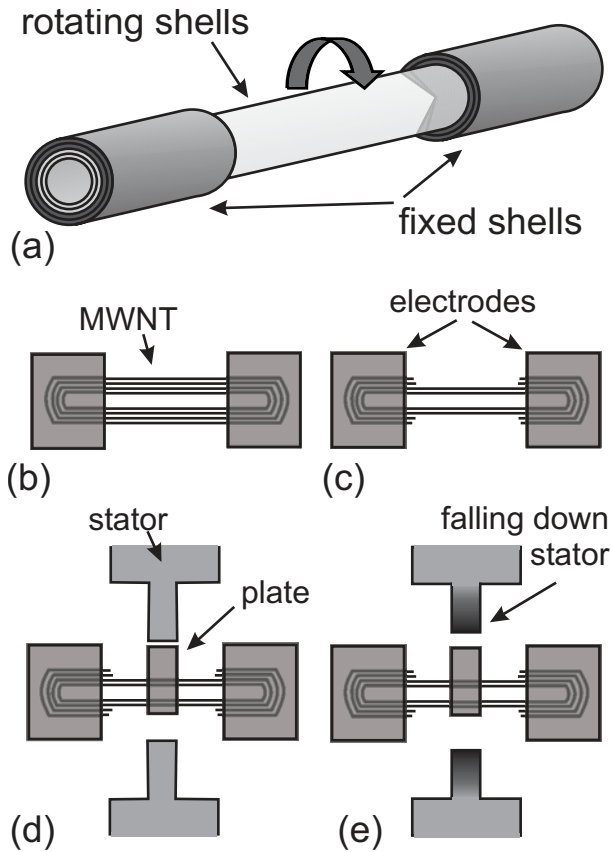}
\caption{(a) Inner shells turn inside fixed outer shells. (b) The
MWNT is contacted to two conducting anchor pads separated by
around \mbox{1 $\mu$m}. (c) Several shells are removed between the
contacts to gain access to a selected inner shell (d) A plate and
two stator electrodes are fabricated. The structures consist of
10nm Cr and 25 nm Au. (e) Etching step with BHF.}
\end{figure}

Fig. 1(d) shows the structures that are fabricated in a second
lithography step. A 500 nm long plate is attached above the
rotating inner shells. The plate is asymmetrically positioned with
respect to the tube so that the longer section can be
electrostatically attracted to one of the two stator electrodes
that are fabricated during the same fabrication step. These 200 nm
wide electrodes are designed to be narrow so that they will be
deposited on the substrate in the etching step making the plate
rotation between the two electrodes easier.

The last step of the process consists of etching 500 nm of the
SiO2 with BHF \cite{Nygard}. The wafer is rinsed in DI and in
ethanol. It is then dried under nitrogen flow. The ethanol is
warmed to just below the boiling point in order to reduce the
surface tension. This reduces the risk of the structures breaking
down during drying.

Figure 2 shows two examples of obtained structures at the end of
the fabrication process. Interestingly, the plate is not
horizontal but has already rotated. The motion has been induced by
surface-tension forces of ethanol when drying the sample. Such
forces are well known in NEMS fabrication to be important and to
deform suspended structures. In contrast, it has been shown that
the plate stays horizontal for samples where outer shells of the
MWNT are not removed \cite{Williams}. This indicates that the
sliding between the moving and the fixed elements of the NEMS
occurs between MWNT shells. Note that the rotation has been
observed on most but not all of the samples. In the cases where
the plate has not moved, some lithography-resist is observed
between the plate and the rest of the structure that may block the
motion.

\begin{figure}
\includegraphics{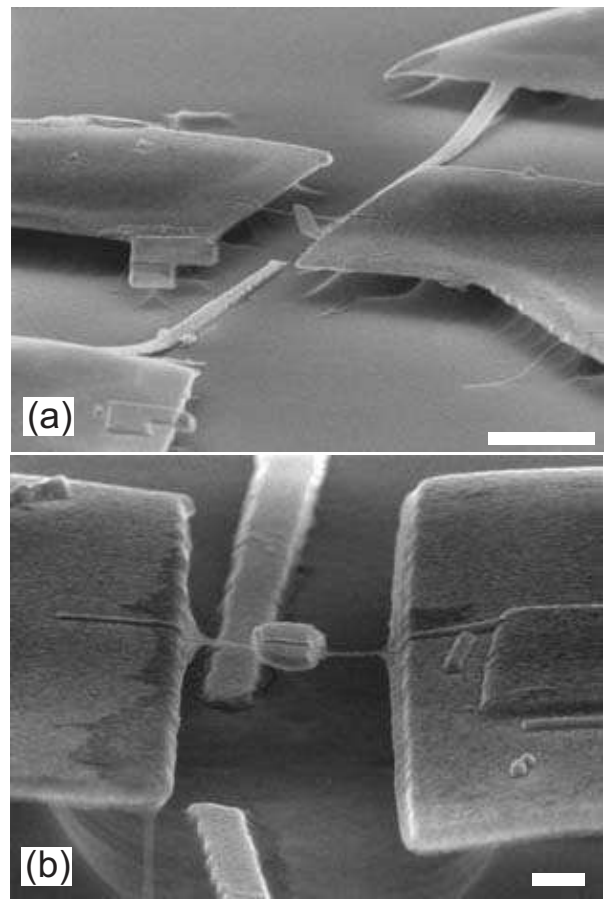}
\caption{SEM images of two samples at the end of the fabrication
process. (a) Scale bar is of length \mbox{1 $\mu$m}. (b) Scale bar
is of length \mbox{200 nm}.}
\end{figure}

We now show that the plate can be electrostatically driven by
stator electrodes. Fig. 3 shows top down images of the device made
with a scanning electron microscope (SEM) that has been modified
in order to electrically access the different electrodes. In Fig.
3(a) the plate is oriented toward the stator electrode on the
right side. A bias voltage $V$ is applied between the plate and
the stator electrode on the left side in order to rotate the plate
in this direction. The motion does not happen at once but occurs
in the following manner. $V$ is increased continuously from zero
while the other stator electrode and the anchor pads are grounded.
The plate stays immobile until 40 V when the plate rotates
suddenly to an almost vertical position (Fig. 3(b)). Another
angular displacement is observed at 49 V (Fig. 3(c)). The bias has
to be increased to 59 V to reach the final position, where the
plate faces the biased electrode (Fig. 3(d)).

Importantly, the plate stays in this direction when the bias is
turned off. This shows that the rotation is not enabled through
the torsion of the tube which would return the plate to the
initial position and which occurs for non engineered MWNTs
\cite{Williams}. We note that the plate stays fixed in this
direction even if 100 V is applied on the right side electrode.
This electrode lies further away from the tube and it is not
surprising that the associated electrostatic torque is not enough
large to initiate the rotation. Optimization of the design
geometry should solve this.

\begin{figure}
\includegraphics{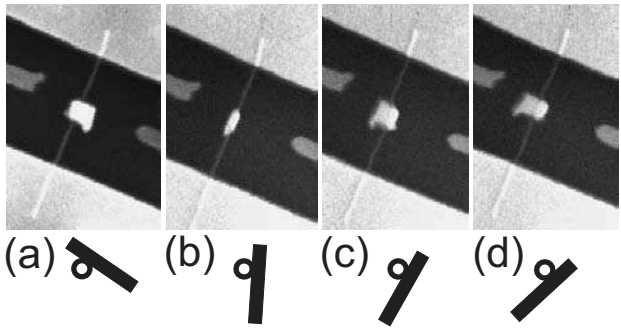}
\caption{SEM images recorded when (a) $V=0$ V (b) $V=40$ V (c)
$V=49$ V (d) $V=59$ V. The cross-section schematics below each SEM
image represent the rotation of the plate around the MWNT. Both
stator electrodes appear less bright than the anchor pads because
they are situated at a lower height.}
\end{figure}

We estimate now the static friction force. Using a finite element
method program and the geometry of the plate and the stator
electrode together with $V=40$ V, we have calculated the
electrostatic energy as a function of the plate position. We
deduce that the electrostatic couple \mbox{$\approx10^{-16}$ Nm}.
This suggests that the static friction force is
\mbox{$\approx2\cdot10^{-15}$ N/\AA$^2$} considering that sliding
happens on a cylindrical surface that is \mbox{1 $\mu$m} long and
18 nm in diameter from AFM. This is a low friction value compared
with previously reported values, which shows that the electrical
breakdown technique is suitable for the fabrication of such NEMSs.

Here we discuss briefly the rotational motion of the plate though
a systematic investigation is left open for further work. The
plate may block at some specific angles due to variations of the
friction between the shells. Friction variation can result from
the displacement of some disorder centers situated along the tube
or some dangling bonds at the extremities left over from the shell
oxidation. Another explanation might be the variation of the
capacitance between the plate and the bias electrode. The blockage
might then occur when the electrostatic torque becomes lower than
the friction torque due to corrugation in the intershell
interaction.

After this work was completed, related results were reported in
Ref. \cite{Fennimore}. A third stator electrode was used in Ref.
\cite{Fennimore} that has allowed for the complete rotation of the
plate around the axle. In contrast to our devices, the
electrical-breakdown technique has not been used to engineer the
MWNT. The plate has been attached directly on the pristine MWNT.
The rotational motion is then obtained by removing one or more
outer shells in the region between the plate and the anchors. The
removal method is to mechanically fatigue and to shear the outer
shells by successive application of large stator voltages.

We wish to acknowledge D. Esteve, P. Joyez, H. Camon and N. Fabre
for discussions, B. Placais, J.M. Berroir and C. Delalande for
support and P. Morfin for technical assitance. LPMC is
CNRS-UMR8551 associated to universities Paris 6 and 7. The
research has been supported by the DGA and ACN programs.

\end{document}